%% file: main.tex
\documentclass[
]{ceurart}

\usepackage{listings}

\newcommand{\arxiv}[2]{#1} 

\usepackage[textsize=scriptsize,tickmarkheight=.1cm]{todonotes}
\usepackage[english]{babel}
\usepackage{amsthm}
\usepackage{mathtools} 
\usepackage{tikz}
\usepackage{physics}
\usepackage{xfrac}
\usepackage{quantikz}
\usepackage[capitalise]{cleveref}
\usepackage{caption}
\usepackage{subcaption}

\setcitestyle{notesep={-}}

\theoremstyle{definition}
\newtheorem{definition}{Definition}[section]
\theoremstyle{example}
\newtheorem{example}{Example}[section]

\newcommand{\neglit}[1]{\lnot #1}
\newcommand{\xor}{\oplus}
\newcommand{\Land}{\bigwedge}
\newcommand{\Lor}{\bigvee}

\newcommand{\tensor}{\otimes}
\renewcommand{\iff}{\leftrightarrow}
\renewcommand{\implies}{\rightarrow}
\newcommand{\VD}{\mathit{VD}} 
\newcommand{\LC}{\mathit{LC}} 
\newcommand{\CZ}{\mathit{CZ}} 
\newcommand{\LCf}{\mathsf{LC}}
\newcommand{\VDf}{\mathsf{VD}}
\newcommand{\EFf}{\mathsf{EF}}
\newcommand{\Idf}{\mathsf{Id}}
\DeclarePairedDelimiter{\ceil}{\lceil}{\rceil}
\newcommand{\neigh}[1]{\mathcal{N}_{#1}}

\DeclareMathSymbol{\shortminus}{\mathbin}{AMSa}{"39}
\newcommand{\mymatrix}{pmatrix}
\tikzset{main/.style={draw,circle}} 
\tikzset{node distance=10mm}

\crefname{definition}{Def.}{Definitions}
\Crefname{definition}{Def.}{Definitions}
\crefname{example}{Example}{Examples}
\Crefname{example}{Example}{Examples}

\DeclareCaptionLabelFormat{figcap1}{\textsf{#1~#2}}
\begin{document}

\copyrightyear{2023}
\copyrightclause{Copyright for this paper by its authors.
  Use permitted under Creative Commons License Attribution 4.0
  International (CC BY 4.0).}

\conference{14th Pragmatics of SAT international workshop}

\title{Quantum Graph-State Synthesis with SAT}


\author[1]{Sebastiaan Brand}[%
orcid=0000-0002-7666-2794,
email=s.o.brand@liacs.leidenuniv.nl,
url=,
]
\cormark[1]

\author[1]{Tim Coopmans}[%
orcid=0000-0002-9780-0949,
email=t.j.coopmans@liacs.leidenuniv.nl,
url=,
]

\author[1]{Alfons Laarman}[%
orcid=0000-0002-2433-4174,
email=a.w.laarman@liacs.leidenuniv.nl,
url=,
]

\address[1]{Leiden Institute of Advanced Computer Science, Leiden University, The Netherlands}

\cortext[1]{Corresponding author.}

\begin{abstract}
	In quantum computing and quantum information processing, graph states are a specific type of quantum states which are commonly used in quantum networking and quantum error correction. A recurring problem is finding a transformation from a given source graph state to a desired target graph state using only local operations. Recently it has been shown that deciding transformability is already NP-hard.
In this paper, we present a CNF encoding for both local and non-local graph state operations, corresponding to one- and two-qubit Clifford gates and single-qubit Pauli measurements. We use this encoding in a bounded-model-checking set-up to synthesize the desired transformation. Additionally, for a completeness threshold on local transformations, we provide an upper bound on the length of the transformation if it exists.
	We evaluate the approach in two settings: the first is the synthesis of the ubiquitous GHZ state from a random graph state where we can vary the number of qubits, while the second is based on a proposed 14 node quantum network. We find that the approach is able to synthesize transformations for graphs up to 17 qubits in under 30 minutes.
\end{abstract}


\maketitle

\input{sections/01-Introduction}
\input{sections/02-Preliminaries}
\input{sections/03-Encoding}

\input{sections/05-Experiments}

\begin{acknowledgments}
This work was supported by the NEASQC project, funded by European Union's Horizon 2020, Grant Agreement No. 951821, and by the Dutch National Growth Fund, as part of the Quantum Delta NL programme.
\end{acknowledgments}

\bibliography{references}

\arxiv{%
\appendix
\input{sections/cnf-appendix}
}{}

\end{document}

%% file: sections/01-Introduction.tex
\section{Introduction}
The creation, manipulation and transmission of quantum information brings into reach applications which are unfeasible or even impossible using classical computers, such as provably-secure communication \cite{bennett1984quantum,ekert1991quantum}, more accurate clock synchronization~\cite{jozsa2000quantum}, and chemistry applications~\cite{cao2019quantum}.
Various questions regarding simulation, modeling and design of quantum computers and networks can be phrased using graph states, a subset of all possible states of a register of quantum bits (qubits) which can be described using graphs~\cite{hein2006entanglement}.
Additionally, graph states are crucial to a universal model of quantum computation called measurement-based quantum computing~\cite{raussendorf2001oneway}.
Furthermore, when augmented with a finite set of quantum operations called Clifford gates and single-qubit Pauli measurements, the graph-state formalism gives rise to efficient classical simulation of a large class of quantum circuits~\cite{anders2006fast} and forms the basis for many quantum error correction schemes~\cite{gottesman1997stabilizer}, a prerequisite for scaling up quantum computing with imperfect devices, as well as many quantum-networking applications~\cite{hillery1999quantum, hein2006entanglement}. 
These applications have a focus on \emph{local} quantum operations, i.e. on a single or few spatially-close qubits, for reasons regarding experimental implementation with imperfect devices.

%



Given this wide applicability, graph-state transformations have been extensively studied from the theory standpoint for various sets of allowed local quantum operations~\cite{nest2004graphical, nest2005thesis, hein2004multiparty,ji2008lulc,hein2006entanglement,englbrecht2022transformations}.
In this work, we consider the following problem: given a source graph state, synthesize a desired target graph state using single-qubit Clifford gates, single-qubit Pauli measurement, and two-qubit Clifford gates between selected pairs of qubits.
When only considering single-qubit Clifford gates and measurements, this problem was shown before~\cite{dahlberg2018transforming} to be equivalent to transforming the associated graphs under two graph operations: an edge-toggling operation called \textit{local complementation} (LC), corresponding to single-qubit Cliffords, and \textit{vertex deletion} (VD), corresponding to measurements. Furthermore, two-qubit Clifford operations can be added to this graph description as \textit{edge flips} (EF) (removing or adding arbitrary edges).
The decision problem ``Can a source graph be transformed to a target graph under LC+VD?'' has been shown to be NP-complete~\cite{dahlberg2020how}, even when restricting the target graph to a practically-relevant scenario~\cite{dahlberg2020transforming}.
Although there exists an algorithm~\cite{dahlberg2018transforming} (based on techniques from~\cite{courcelle2007vertex, courcelle2012graph}) which is fixed-parameter tractable (FPT) in the rank-width $r$ of the graph, the authors of the algorithm themselves remark it is not useful in practice due to a giant FPT-prefactor equalling ten times repeated exponentiation with base 2 (i.e. $2^{2^{\dots 2^r}}$)~\cite{dahlberg2020how}.
Additionally, for determining LC+VD transformability, Dahlberg et al.~\cite{dahlberg2018transforming} reduce the search space based on a necessary and sufficient condition. The size of the search space however remains exponential in the number of nodes $n$, and their brute-force algorithm for traversing the space runs in $O(3^{n-k} \cdot \mathsf{poly}(n,k))$ time, where $k$ is the number of required vertex deletions.

We tackle the problem of graph-state synthesis under LC+VD+EF with bounded model checking (BMC)~\cite{clarke2001bounded,biere2009bounded}. To this end we present a Boolean encoding for graph states and the operations on them, and provide a completeness threshold for reachability under LC+VD.
This approach can be applied to arbitrary graphs, in contrast to special cases for which poly-time algorithms have been found~\cite{dahlberg2020how,hahn2019quantum} or unsatisfiability can be determined analytically~\cite{hahn2022limitations,jong2023extracting}.
We evaluate this approach in two settings of particular interest~\cite{dahlberg2020how,jong2023extracting}: first, we synthesize the ubiquitous Greenberger–Horne–Zeilinger (GHZ) state~\cite{greenberger1989going} from random graphs with varying number of qubits. Next, we target a 14 node quantum network proposal~\cite{rabbie2022designing}.
We find that when searching for transformations under LC+VD+EF, where the pairs of nodes between which edge flips are allowed to take place can be selected, the BMC approach finds transformations for graphs up to 17 nodes (qubits) within 30 minutes.
In comparison, for transformations under single-qubit Clifford operations without measurements (a setting where deciding reachability is in P~\cite{bouchet1991efficient,nest2004efficient} and counting reachable graphs is \#P-complete~\cite{dahlberg2020counting}), various properties of equivalence classes have been explored up to 12 qubits~\cite{danielsen2006classification,cabello2011optimal,adcock2020mapping}.
Our approach can also find transformations under only LC+VD, although in this setting it is not faster than the brute-force algorithm from~\cite{dahlberg2018transforming}.

Aside from graph problems which have been tackled with SAT-based methods~\cite{heule2015sat,samer2009encoding,schidler2022sat,ganian2019sat,lodha2017sat,lodha2016sat,lodha2019sat,courcelle2022using},
SAT has also been used on problems in quantum computing.
For example, synthesizing optimal Clifford circuits without measurements (closely related to graph-state synthesis under LC + flipping arbitrary edges, but without VD) has been tackled with BMC~\cite{schneider2023sat}.
Without the optimality constraint (i.e. shortest circuit) this problem is in P~\cite{aaronson2004improved}, while the complexity with the optimality constraint is unknown.
SAT-based techniques have also been applied to quantum circuit equivalence checking for a limited selection of circuits~\cite{yamashita2010fast}.
BMC specifically has been applied to Clifford circuit (without measurements) equivalence checking~\cite{berent2022towards} (a problem that is also in P~\cite{amy2018towards}), and SMT and planning based approaches have been used to map logical quantum circuits to physical quantum-chips~\cite{tan2020optimal,shaik2023optimal}. 
Unlike much previous work we include measurements, which for our problem raises the complexity from P to NP-complete.

%% file: sections/02-Preliminaries.tex
\section{Preliminaries and problem definition}

\subsection{Quantum computing}

We very briefly introduce quantum bits (qubits) and how to act on them with quantum gates and measurements (see \cite{nielsen2002quantum} for a complete introduction).
The state $\ket{\psi}$ of a single qubit is a complex 2-vector of unit norm, equalling the \emph{computational-basis states} $\ket{0} = \begin{\mymatrix}1 & 0\end{\mymatrix}^\intercal$ or $\ket{1} = \begin{\mymatrix}0 & 1\end{\mymatrix}^\intercal$ or any linear combination of those, i.e. in general a single-qubit state is $\ket{\psi} = \alpha_0 \ket{0} + \alpha_1 \ket{1} = \begin{\mymatrix} \alpha_0 & \alpha_1\end{\mymatrix}^\intercal$ for complex numbers $\alpha_0, \alpha_1$ satisfying $|\alpha_0|^2 + |\alpha_1|^2 = 1$ (here, $\intercal$ denotes vector transposition).
	A general $n$-qubit quantum state is represented as a complex vector of length $2^n$ with norm 1, e.g. \mbox{$\begin{\mymatrix} \frac{1}{\sqrt{2}} & \frac{i}{\sqrt{2}} \end{\mymatrix}^{\intercal}$ and $\begin{\mymatrix} \frac{2}{\sqrt{13}} & 0 & 0 & \shortminus\frac{3}{\sqrt{13}} \end{\mymatrix}^{\intercal}$} are quantum states.
The joint state of two separate quantum registers in states $\ket{\phi}, \ket{\psi}$ is $\ket{\phi}\otimes \ket{\psi}$, where $\otimes$ denotes the tensor product: given $r_V \times c_V$ matrix $V$ and $r_W \times c_W$ matrix $W$, the $r_V r_W \times c_V c_W$ matrix $V \otimes W$ is
\[
V \tensor W = 
\begin{\mymatrix}
V_{00} W & V_{01}W &\dots& V_{0 c_V} W\\
\vdots&\vdots&\ddots&\\
V_{r_V 0} W & V_{r_V 1}W &\dots& V_{r_V c_V} W\\
\end{\mymatrix}
.
\]
Given a bipartition $A \cup B = \{1, 2, \dots, n\}$, an $n$-qubit state $\ket{\psi}$ is called \emph{separable over $A,B$} if we can write $\ket{\psi} = \ket{\varphi}_A \tensor \ket{\phi}_B$.
It is \emph{entangled} otherwise, a feature that has no classical analogue and is a prerequisite to many applications with a quantum advantage. For example $\begin{\mymatrix} \frac{1}{\sqrt{2}} & 0 & \frac{1}{\sqrt{2}} & 0\end{\mymatrix}^{\intercal} = 
\begin{\mymatrix} \frac{1}{\sqrt{2}} & \frac{1}{\sqrt{2}} \end{\mymatrix}^{\intercal} 
	\otimes
\begin{\mymatrix} 1 & 0 \end{\mymatrix}^{\intercal}$ is not entangled, but 
	$
\begin{\mymatrix} \frac{1}{\sqrt{2}} & 0 & 0 & \frac{1}{\sqrt{2}} \end{\mymatrix}^{\intercal} 
	$ is.



\begin{wrapfigure}{r}{4.1cm}
\vspace{-.5cm}
\scalebox{.9}{\parbox{.5\linewidth}{%
\begin{equation}
\nonumber
\begin{gathered}
I = 
\begin{\mymatrix}
1 & 0 \\ 0 & 1
\end{\mymatrix}
~~T =
\begin{\mymatrix}
1 & 0 \\ 0 & e^{i\pi/4}
\end{\mymatrix}
\\
H =
\frac{1}{\sqrt{2}}
\begin{\mymatrix}
1 & 1 \\ 1 & \shortminus 1
\end{\mymatrix}
\\
\CZ =
\begin{\mymatrix}
1 & 0 & 0 & 0 \\
0 & 1 & 0 & 0 \\
0 & 0 & 1 & 0 \\
0 & 0 & 0 & \shortminus 1 
\end{\mymatrix}
\end{gathered}
\end{equation}
}}
\vspace{-1.2cm}
\end{wrapfigure}

A quantum gate (always reversible) on $n$ qubits is given by a $2^n \times 2^n$ unitary matrix and the output state can be found by matrix-vector multiplication, for example $H$ (see right) which maps input $\begin{\mymatrix}1 \\ 0\end{\mymatrix}$ to output $\frac{1}{\sqrt{2}} \begin{\mymatrix}1 & 1\\ 1 & \shortminus 1 \end{\mymatrix}\cdot \begin{\mymatrix}1 \\ 0\end{\mymatrix} = \begin{\mymatrix}\frac{1}{\sqrt{2}} \\ \frac{1}{\sqrt{2}} \end{\mymatrix}$.
An example universal gate set is shown on the right.
The tensor product is used to apply gates in parallel to separate registers, e.g.
 $I \tensor H \tensor I$ is a 3-qubit gate performing a $H$ on the second qubit and $I$ on the first and third.
The result of a two-qubit gate (e.g. \mbox{controlled-$Z$} ($\CZ$), which maps e.g.
\mbox{
$\begin{\mymatrix} \frac{1}{\sqrt{2}} & 0 & 0 & \frac{1}{\sqrt{2}} \end{\mymatrix}^{\intercal} $
	to
$\begin{\mymatrix} \frac{1}{\sqrt{2}} & 0 & 0 & \shortminus\frac{1}{\sqrt{2}} \end{\mymatrix}^{\intercal} $
}) 
between two non-adjacent qubits can be computed by swapping qubits: 
e.g. for qubits $q_0, q_1, q_2$, $\CZ(q_0,q_2) = \text{SWAP}(q_1,q_2)\CZ(q_0,q_1)\text{SWAP}(q_1,q_2)$, where $\text{SWAP}(q_1,q_2)$ replaces $\ket{a}\otimes\ket{b}\otimes\ket{c} \rightarrow \ket{a}\otimes\ket{c}\otimes\ket{b}$ for $a,b,c \in\{0, 1\}$.
The gates $H, T^2$ together generate (under matrix multiplication and tensoring with $I$) the group of \emph{single-qubit Clifford gates}, and $H, T^2, \CZ$ together generate all Clifford gates.

A computational-basis measurement is a non-reversible operation which projects a single qubit state $\alpha_0 \ket{0} + \alpha_1 \ket{1}$ to one of $\ket{0},\ket{1}$ with probability $|\alpha_0|^2$ or $|\alpha_1|^2$. For example, measuring a qubit $\ket{\psi} = \sqrt{\sfrac{1}{3}}\ket{0} + \sqrt{\sfrac{2}{3}}\ket{1}$ yields the state $\ket{0}$ with probability $1/3$ and the state $\ket{1}$ with probability $2/3$. Any $n$-qubit state $\ket{\psi}$ can be written as $\ket{\psi} = \alpha\ket{0} \tensor \ket{\psi_0} + \beta\ket{1} \tensor \ket{\psi_1}$ where $|\alpha|^2$ ($|\beta|^2$) is the probability of finding the first qubit in the $\ket{0}$ ($\ket{1}$) state after measuring it (for expressing measurement on the other qubits, swap qubits first).
A Pauli measurement equals a computational-basis measurement preceded by a single-qubit Clifford gate.
Sequences of quantum operations are typically visualized in a quantum circuit (see \cref{fig:circuit}). 
\vspace{-0.5\baselineskip}

\begin{figure}[t]
\begin{minipage}{.4\textwidth}
\centering
\begin{quantikz}[row sep={0.1cm}]
\lstick{$\ket{0}$}\slice{$\ket{\psi_1}$} & \gate{H}\slice{$\ket{\psi_2}$} & \ctrl{1}\slice{$\ket{\psi_3}$} & \qw\slice{$\ket{\psi_4}$} & \meter{} \\
\lstick{$\ket{0}$} & \gate{H} & \control[]{} & \gate{H} & \meter{}
\end{quantikz}
\end{minipage}%
\begin{minipage}{.6\textwidth}
\begin{align*}
\ket{\psi_1} =&~ \ket{0} \tensor \ket{0} = \ket{00} \\
\ket{\psi_2} =&~ (H \tensor H) \ket{\psi_1} = \tfrac{1}{2} (\ket{00} + \ket{01} + \ket{10} + \ket{10}) \\
\ket{\psi_3} =&~ \CZ\ket{\psi_2} = \tfrac{1}{2}  (\ket{00} + \ket{01} + \ket{10} - \ket{11}) \\
\ket{\psi_4} =&~ (I \tensor H) \ket{\psi_3} = \tfrac{1}{\sqrt{2}}(\ket{00} + \ket{11})
\end{align*}
\end{minipage}
\caption{An example 2-qubit quantum circuit.
Operations are applied from left to right. The \mbox{controlled-$Z$} ($\CZ$) gate is visualized as
{\protect\tikz[baseline=.1ex]{
\protect\fill (0,0) circle (1.5pt) coordinate (A);
\protect\fill (0,1.5ex) circle (1.5pt) coordinate (B);
\protect\draw [line width=.2mm] (A)--(B);}
\hspace{-.2cm}
}
. 
As is common, we write $\ket{01}$ as shorthand for $\ket{0}\otimes\ket{1}$, $\ket{01} = \ket{0}\otimes\ket{1}$, etc.
Measuring both qubits at the end gives $\ket{00}$ or $\ket{11}$ with equal probability.
}
\label{fig:circuit}
\end{figure}

\subsection{Graph states and graph-state reachability}
\label{sec:prelim-graph-states}
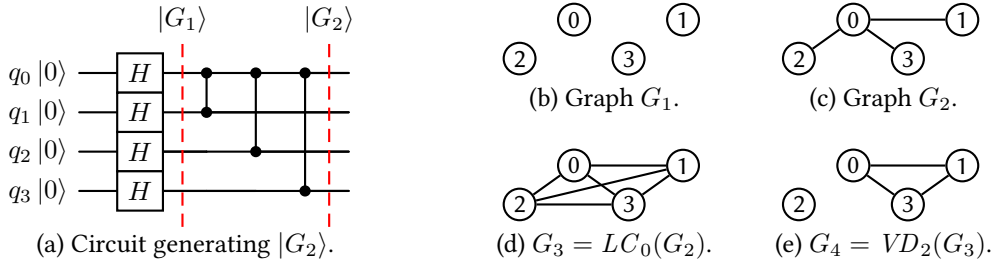
\begin{figure}[t]
\begin{minipage}[b]{0.5\textwidth}
\begin{subfigure}[b]{\textwidth}
	\centering
	\begin{quantikz}[row sep={0cm}]
	\lstick{$q_0 \ket{0}$} & \gate{H}\slice{$\ket{G_1}$} & \ctrl{1} & \ctrl{2} & \ctrl{3}\slice{$\ket{G_2}$} & \qw \\
	\lstick{$q_1 \ket{0}$} & \gate{H} & \ctrl{0} & \qw & \qw & \qw \\
	\lstick{$q_2 \ket{0}$} & \gate{H} & \qw & \ctrl{0} & \qw & \qw \\
	\lstick{$q_3 \ket{0}$} & \gate{H} & \qw & \qw & \ctrl{0} & \qw
	\end{quantikz}
	\vspace{-.2cm}
	\caption{Circuit generating $\ket{G_2}$.}
	\label{fig:example-circuit}
\end{subfigure}
\end{minipage}%
\begin{minipage}[b]{0.5\textwidth}
\begin{subfigure}[b]{0.5\textwidth}
	\centering
	\begin{tikzpicture}[auto, thick,node distance=.4cm,inner sep=1.5pt]
		\node[draw,circle] (0) [] {0};
		\node[draw,circle] (2) [below left = .2cm and .4cm of 0] {2};
		\node[draw,circle] (3) [below right = .2cm and .4cm of 0] {3};
		\node[draw,circle] (1) [above right = .2cm and .4cm of 3] {1};
	\end{tikzpicture}
	\vspace{-.1cm}
	\caption{Graph $G_1$.}
	\label{fig:example-graph1}
\end{subfigure}%
\begin{subfigure}[b]{0.5\textwidth}
	\centering
	\begin{tikzpicture}[auto, thick,node distance=.4cm,inner sep=1.5pt]
		\node[draw,circle] (0) [] {0};
		\node[draw,circle] (2) [below left = .2cm and .4cm of 0] {2};
		\node[draw,circle] (3) [below right = .2cm and .4cm of 0] {3};
		\node[draw,circle] (1) [above right = .2cm and .4cm of 3] {1};
		\draw[-] (0) to (1);
		\draw[-] (0) to (2);
		\draw[-] (0) to (3);
	\end{tikzpicture}
	\vspace{-.1cm}
	\caption{Graph $G_2$.}
	\label{fig:example-graph2}
\end{subfigure}\vspace{.4cm}
\begin{subfigure}[b]{0.5\textwidth}
	\centering
	\begin{tikzpicture}[auto, thick,node distance=.4cm,inner sep=1.5pt]
		\node[draw,circle] (0) [] {0};
		\node[draw,circle] (2) [below left = .2cm and .4cm of 0] {2};
		\node[draw,circle] (3) [below right = .2cm and .4cm of 0] {3};
		\node[draw,circle] (1) [above right = .2cm and .4cm of 3] {1};
		\draw[-] (0) to (1);
		\draw[-] (0) to (2);
		\draw[-] (0) to (3);
		\draw[-] (1) to (2);
		\draw[-] (2) to (3);
		\draw[-] (1) to (3);
	\end{tikzpicture}
	\vspace{-.1cm}
	\caption{$G_3 = \LC_0(G_2)$.}
	\label{fig:lc-example}
\end{subfigure}%
\begin{subfigure}[b]{0.5\textwidth}
	\centering
	\begin{tikzpicture}[auto, thick,node distance=.4cm,inner sep=1.5pt]
		\node[draw,circle] (0) [] {0};
		\node[draw,circle] (2) [below left = .2cm and .4cm of 0] {2};
		\node[draw,circle] (3) [below right = .2cm and .4cm of 0] {3};
		\node[draw,circle] (1) [above right = .2cm and .4cm of 3] {1};
		\draw[-] (0) to (1);
		\draw[-] (0) to (3);
		\draw[-] (1) to (3);
	\end{tikzpicture}
	\vspace{-.1cm}
	\caption{$G_4 = \VD_2(G_3)$.}
	\label{fig:vd-example}
\end{subfigure}
\end{minipage}
\caption{The circuit in \ref{fig:example-circuit} generates the state $\ket{G_2}$, corresponding to the graph in \ref{fig:example-graph2}. Examples of local complementation and vertex deletion are shown in \ref{fig:lc-example} and \ref{fig:vd-example}.}
\label{fig:gs-examples}
\vspace{-.1cm}
\end{figure}

Graph states are a subset of all quantum states.
An $n$-qubit graph state $\ket{G}$ is represented by an undirected simple graph $G = (V, E)$ with $|V| = n$ vertices and no self-loops (where $V$ is the vertex set and $E \subseteq V \times V$ the edge set), constructed as starting from the state $H^{\otimes n}\ket{0}^{\otimes n}$, followed by a $\CZ$ gate on each pair of qubits $(u, v) \in E$. An example is given in \cref{fig:gs-examples}.
From here on we say `graph' to mean `undirected simple graph without self-loops'.
Intuitively, the graph $G$ captures information about the entanglement between the qubits, where two qubits are entangled if they are (directly or indirectly) connected in the graph.


The two graph transformations corresponding to single-qubit quantum operations are:
\begin{itemize}
\item{
\emph{Local complementation} $\LC_k$ on vertex $k \in V$ transforms $G = (V,E)$ into $LC_k(G)=(V,E')$ where $E'$ is obtained from $E$ by flipping the edges in the neighborhood of $k$, i.e. for all $u, v \in \neigh k$, if $(u,v) \in E$ then $(u,v) \not\in E'$ and if $(u,v) \not\in E$ then $(u,v) \in E'$.
Here, the neighborhood $\neigh k$ is the set of all vertices adjacent to $k$, i.e. $\neigh k = \{v \mid (k,v) \in E\}$.
For any graphs $G$ and $G'$, $\ket{G'}$ is reachable from $\ket{G}$ using only single-qubit Clifford operations if and only if $G'$ is reachable from $G$ using local complementations.
More specifically, the graph state $\ket{\LC_k(G)}$ equals the resulting quantum state when applying a certain sequence of single-qubit Clifford operations to $\ket{G}$ (see \cite{nest2004graphical} for details).
}
\item{
\emph{Vertex deletion} $\VD_k$ of vertex $k \in V$ transforms $G = (V, E)$ to $\VD_k(G) = (V, E')$ with $E' = E \setminus \{ (v, k) \mid v \in V\}$, i.e. $k$ becomes \emph{isolated} (all edges adjacent to $k$ are removed).
		Vertex deletion of vertex $k$ implements measurement on qubit $k$: for each graph $G$, the graph state $\ket{\VD_k(G)}$ is single-qubit Clifford equivalent to $\ket{G}$ at which a computational-basis measurement has been performed on qubit $k$~\cite{hein2004multiparty}.
}
\end{itemize}
And although not the primary focus of this work, we can also consider two-qubit operations:
\begin{itemize}
\item{
Given a subset of pairs of nodes $D \subseteq V \times V$ (for convenience $u < v$ for $(u,v) \in D$), $G$ can be transformed into $G'$ by \emph{edge flips} among $D$ and local complementations on vertices in $V$ if and only if $\ket{G}$ can be transformed into $\ket{G'}$ using two-qubit Clifford operations on the qubit pairs in $D$ and single-qubit Clifford on qubits in $V$~\cite[Th.1]{englbrecht2022transformations}.
}
\end{itemize}

\label{sec:gs-reach}



Rather than generating a graph state from scratch using $\CZ$ gates (as in \cref{fig:example-circuit,fig:example-graph1,fig:example-graph2}) a problem of interest for e.g. quantum networking is to obtain a particular graph from an existing graph \emph{using only single-qubit operations} (LC+VD, and $D=\emptyset$). Below is a practical example.

\vspace{0.5\baselineskip}
\noindent
\begin{minipage}[b]{.8\textwidth}
\begin{example}
\label{ex:secret-sharing}
Alice is part of a 6-node quantum network and wants to run a quantum secret sharing scheme~\cite{hillery1999quantum} between herself and three other parties, each having one qubit. For this she needs a 4-qubit Greenberger–Horne–Zeilinger (GHZ) state~\cite{greenberger1989going}, given by $G_{\text{GHZ}}$ on the right.
However, generating $\ket{G_{\text{GHZ}}}$ using $\CZ$-gates (\cref{fig:example-circuit,fig:example-graph1,fig:example-graph2})) requires generating entanglement~\cite[Fig.2.4]{eisert2000optimal,nickerson2015practical}, 
which is a time-consuming probabilistic process~\cite{sangouard2011quantum}.
At some point in time the network is a state $\ket{G_s}$.
Because single-qubit operations (LC+VD on the graph) are much easier to perform than entanglement generation,
Alice wants to know whether a given $G_s$ can be transformed into $G_{\text{GHZ}}$ using only LC+VD.
\end{example}
\end{minipage}
\begin{minipage}[b]{.19\textwidth}
\centering
\scalebox{.9}{%
\begin{tikzpicture}[auto,thick,node distance=.4cm,inner sep=1.5pt]
	\node[draw,circle, minimum size=5mm] (0) [] {};
	\node[draw,circle, minimum size=5mm] (2) [below left = .2cm and .3cm of 0] {B};
	\node[draw,circle, minimum size=5mm] (3) [below right = .2cm and .3cm of 0] {A};
	\node[draw,circle, minimum size=5mm] (1) [above right = .2cm and .3cm of 3] {C};
	\node[draw,circle, minimum size=5mm] (4) [below right = .2cm and .3cm of 2] {};
	\node[draw,circle, minimum size=5mm] (5) [below right = .2cm and .3cm of 3] {D};
	\draw[-] (3) to (0);
	\draw[-] (0) to (2);
	\draw[-] (3) to (1);
	\draw[-] (3) to (4);
	\draw[-] (4) to (5);
	\node[] (caption) [above right = 0cm and .3cm of 0] {$G_s$};
\end{tikzpicture}}\\\vspace{.3cm}
\scalebox{.9}{%
\begin{tikzpicture}[auto, thick,node distance=.4cm,inner sep=1.5pt]
	\node[draw,circle, minimum size=5mm] (0) [] {};
	\node[draw,circle, minimum size=5mm] (2) [below left = .2cm and .3cm of 0] {B};
	\node[draw,circle, minimum size=5mm] (3) [below right = .2cm and .3cm of 0] {A};
	\node[draw,circle, minimum size=5mm] (1) [above right = .2cm and .3cm of 3] {C};
	\node[draw,circle, minimum size=5mm] (4) [below right = .2cm and .3cm of 2] {};
	\node[draw,circle, minimum size=5mm] (5) [below right = .2cm and .3cm of 3] {D};
	\draw[-] (3) to (2);
	\draw[-] (3) to (1);
	\draw[-] (3) to (5);
	\node[] (caption) [above right = 0cm and 0cm of 0] {$G_{\text{GHZ}}$};
\end{tikzpicture}
}
\end{minipage}

\vspace{0.5\baselineskip}

This motivates the problem we will study in this work, posed before in~\cite{dahlberg2020how} for single-qubit operations (LC+VD and $D=\emptyset$) and in~\cite{englbrecht2022transformations} for multi-qubit operations (LC+VD and $D\neq \emptyset$).

\begin{definition}[Graph-state synthesis]
\label{def:gs-reach}
	Given source and target graphs $G_s = (V,E_s)$ and $G_t = (V,E_t)$, find (if it exists) a sequence of local complementations and vertex deletions on any $v \in V$ (and also edge flips on $(u,v) \in D$ for some given $D \subseteq V \times V$ in case multi-qubit Clifford operations are allowed on $D$) which transforms $G_s$ into $G_t$.
\end{definition}

We remark that if $D= V\times V$, any graph can be trivially synthesized because an edge may be added or removed between any pair of nodes (\cref{fig:example-circuit,fig:example-graph1,fig:example-graph2}).
We also remark that we are not necessarily interested in the shortest sequence of graph transformations, as any sequence of LC+VD translates into at most one single-qubit Clifford and one measurement per qubit.

%% file: sections/03-Encoding.tex
\newcommand{\vv}{k} 
\newcommand{\vA}{u} 
\newcommand{\vB}{v} 

\section{SAT encoding}
\label{sec:encoding}
As seen in the previous section, quantum operations on graph states can be expressed through graph transformations. In this section, we give Boolean encodings for these operations, as well as an encoding for the transition relation as a whole.
\arxiv{In \cref{app:cnf} we detail how these Boolean expressions are written in conjunctive normal form (CNF).

}{}
The encoding of a single transformation step from graph $G$ to $G'$ uses variables $\vec x$ for $G$ and $\vec x'$ for $G'$.
We encode a graph as follows.


\begin{definition}[Graph encoding]
\label{def:graph-enc}
An undirected graph $G$ of $n$ vertices is encoded as a conjunction over $n(n-1)/2$ literals $x_{\vA\vB}$ ($\neglit{x_{\vA\vB}}$), for $(u,v) \in \mathbb U = \{(u,v) \in V \times V \mid u < v\}$, indicating there is (not) an edge between nodes $\vA$ and $\vB$.
\end{definition}

\subsection{Encoding of graph transformations}
The Boolean encoding for deleting a vertex $\vv$, denoted $\VDf_\vv$, is given in \cref{eq:vd-k}. All edges $(\vA,\vB)$ connected to $\vv$ are set to false ($\neglit x_{\vA\vB}'$) while all others remain unchanged ($x_{\vA\vB}' \iff x_{\vA\vB}$).
\begin{equation}
\VDf_\vv = \Land_{(\vA,\vB) \in \mathbb U} 
\begin{cases}
\neglit{x_{\vA\vB}'} & \textnormal{ if $\vA = \vv$ or $\vB = \vv$} \\
x_{\vA\vB}' \iff x_{\vA\vB} & \textnormal{ otherwise.}
\end{cases}
\label{eq:vd-k}
\end{equation}

The encoding for performing a local complementation on vertex $\vv$, denoted $\LCf_\vv$, is given in \cref{eq:lc-k} and can be read as follows: if vertices $\vA,\vB$ are in the neighborhood of $\vv$ ($x_{\vA\vv} \land x_{\vB\vv}$) then the value of the edge $(\vA,\vB)$ is flipped ($x_{\vA\vB}' \iff \lnot (1 \xor \neglit x_{\vA\vB})$.
\begin{equation}
\LCf_\vv = \Land_{(\vA,\vB) \in \mathbb U}
\begin{cases}
x_{\vA\vB}' \iff \lnot (( x_{\vA\vv} \land x_{\vB\vv} ) \xor \neglit x_{uv}) & \textnormal{ if $\vA \neq \vv$ and $\vB \neq \vv$} \\
x_{\vA\vB}' \iff x_{\vA\vB}& \textnormal{ otherwise.}
\end{cases}
\label{eq:lc-k}
\end{equation}


To encode edge flips on a selection of edges $D$ (\cref{def:gs-reach}), we take $D$ to be an indexed set $D = \{(u_1, v_1), (u_2, v_2), \dots \}$ with $u_i < v_i$.
Given this indexed set, the constraint in \cref{eq:ef-i} encodes an edge flip of $(\vA_i,\vB_i)$.
\begin{equation}
\EFf_i = \Land_{(\vA,\vB) \in \mathbb U}
\begin{cases}
x_{\vA\vB}' \xor x_{\vA\vB} &\text{if $u = u_i$ and $v = v_i$} \\
x_{\vA\vB}' \iff x_{\vA\vB} &\text{otherwise}
\end{cases}
\label{eq:ef-i}
\end{equation}

In order to combine the transition relations $\LCf_\vv$, $\VDf_\vv$, and $\EFf_i$ into a single CNF formula we use a construction similar to the BMC encoding of different concurrent threads in~\cite{rabinovitz2005bounded}: we add $\ceil{\log_2 (\max(|V|,|D|) + 1)}$ variables $\vec y$ for the binary encoding of $\vv \in V$ or $i \in \{1, \dots, |D|\}$, and two variables $\vec z$ to indicate whether a given operation is a local complementation ($\vec z = 0$), a vertex deletion ($\vec z = 1$), or an edge flip ($\vec z = 2$). For example the constraint $\vec y = 3 \land \vec z = 1$ represents vertex deletion of node $3$.
Using these additional variables, we encode all local complementations, vertex deletions, and edge flips as in \cref{eq:all-lc,eq:all-vd,eq:all-ef}.

\begin{equation}
\label{eq:all-lc}
R_\LCf(\vec x, \vec x') = \Land_{\vv \in V}
\Big[ ( \vec y = \vv \land \vec z = 0 ) \implies \LCf_\vv (\vec x, \vec x') \Big]
\end{equation}

\begin{equation}
\label{eq:all-vd}
R_\VDf (\vec x, \vec x') = \Land_{\vv \in V} 
\Big[ ( \vec y = \vv \land \vec z = 1 ) \implies \VDf_\vv (\vec x, \vec x') \Big]
\end{equation}

\begin{equation}
\label{eq:all-ef}
R_\EFf (\vec x, \vec x') = \Land_{i \in \{1, \dots, |D|\}} 
\Big[ (\vec y = i \land \vec z = 2) \implies \EFf_i(\vec x, \vec x') \Big]
\end{equation}


Additionally we add an identity transition $R_\Idf (\vec x, \vec x') = ( \vec z = 3 ) \implies \Idf (\vec x, \vec x')$ to ensure that if a transformation of length $d$ exists, a transformation of length $d' \geq d$ also exists (to avoid searching over all $d$), and we appropriately constrain the unused values of $\vec y$ and $\vec z$ by adding $C = (\vec{y}~<~|V|~\lor~z~=~2) \land (\vec{y}~<~|D| \lor z \neq 2)$. Finally, we obtain the the global transition relation in \cref{eq:R-global}. 
When converted to CNF this formula has $m + n(n-1)$ variables and $\leq 3.5n^3 + 2mn^2 + 0.5n^2 + 0.5|D|n^2$ clauses, where $n=|V|$ and $m = \ceil{\log_2 (\max(|V|,|D|) + 1)}$.


\begin{equation}
\label{eq:R-global}
R_{\text{global}}(\vec x, \vec x') = R_{\LCf} \land R_{\VDf} \land R_{\EFf} \land R_\Idf \land C
\end{equation}

We use the transition relation specified in \cref{eq:R-global} in a bounded-model-checking set-up, i.e. we create \cref{eq:gs-bmc} below, where $S(\vec x_1)$ encodes a source graph $G_s$, $T(\vec x_d)$ a target graph $G_t$, and $d$ is the search depth.
\begin{equation}
S(\vec x_1) \land \Land_{i=1}^{d-1} R_{\text{global}}(\vec x_{i}, \vec x_{i+1}) \land T(\vec x_d)
\label{eq:gs-bmc}
\end{equation}
The formula is satisfiable if and only if a sequence of operations of at most $d$ steps exists which transforms $G_s$ into $G_t$. In \cref{sec:max-depth}, we prove an upper bound on the required depth $d$.

\vspace{-0.5\baselineskip}
\subsection{Completeness threshold}
\label{sec:max-depth}
To provide a completeness threshold for graph-state synthesis under LC+VD, we use the following observations to bound the search depth.
\begin{enumerate}
	\item If $G_s$ can be transformed to $G_s'$ under LC, a transformation exists of at most $M$ local complementations, where $M = 3(|V|-s)/2$ with $s=|V| (\textnormal{mod } 2$)~\cite[\S 4]{bouchet1991efficient}.
\item If $G_s$ can be transformed into $G_t$ under LC+VD, then vertex deletion needs to be performed on exactly the $\Delta$ vertices which are isolated in $G_t$.\footnote{Without loss of generality, we assume $G_s$ has no isolated vertices. If $G_s$ has isolated vertices which are not isolated in $G_t$, then $G_t$ is trivially unreachable under LC+VD.} 
\item For $k \in V$, $LC_k$ after $\VD_k$  leaves the graph unchanged, i.e. $\LC_k(\VD_k(G)) = \VD_k(G)$.
\item For $j,k\in V$ and $j \neq k$, $LC_j$ and $\VD_k$ commute, i.e. $\LC_j(\VD_k(G)) = \VD_k(LC_j(G))$.
\end{enumerate}

\begin{wrapfigure}{r}{6.5cm}
\centering
\vspace{-.5cm}
\begin{quantikz}[row sep={0.1cm}, column sep={0.5cm},wire types={b,b}, classical
gap=0.12cm]
\lstick[wires=2]{$|V|$} \slice{$\ket{G_s}$}
& \gate[2]{SQC} \slice{$\ket{G_s'}$}
& \meter{} \slice{$\ket{G_t}$} & \\ 
& & & 
\rstick{\hspace{-1mm}$\}~|V|-\Delta$}
\end{quantikz}
\vspace{-.3cm}
\captionsetup{width=6.5cm}
\caption{Graph-state transformation circuit under LC+VD.}
\label{fig:gs-transform-circ}
\end{wrapfigure}
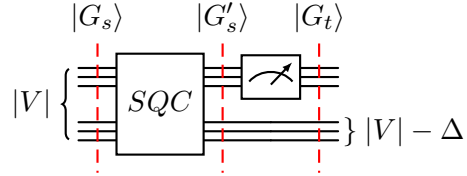

\noindent
From points 3 and 4, it follows that all vertex deletions (measurements) can be postponed until after the local complementations (single-qubit Clifford gates). We then get that if $G_s$ can be transformed into $G_t$ under LC+VD, it can be transformed by a circuit of the form given in \cref{fig:gs-transform-circ}, taking at most $M$ local complementations and $\Delta$ vertex deletions.

%% file: sections/05-Experiments.tex
\section{Empirical evaluation}
We evaluate our approach in two settings: synthesizing a GHZ state from random graphs for an increasing number of qubits, and synthesis of graphs based on a proposal of a 14 node quantum network in the Netherlands~\cite{rabbie2022designing}. 
For all experiments, we perform binary search over $d$ up to the completeness threshold specified in \cref{sec:max-depth}. In our current setup, the solver is restarted for every different $d$.
Experiments\footnote{Reproducible experiments are available online at \url{https://github.com/sebastiaanbrand/graph-state-synthesis}.} were run on Ubuntu 18 with an AMD Ryzen 7 5800x CPU. Two different SAT solvers, Glucose~4~\cite{audemard2018glucose} and Kissat~\cite{biere2022kissat}, have been used.

\begin{figure}[t]
\centering
\hspace{-.2cm}
\begin{subfigure}[b]{0.7\textwidth}
\includegraphics[height=4.0cm]{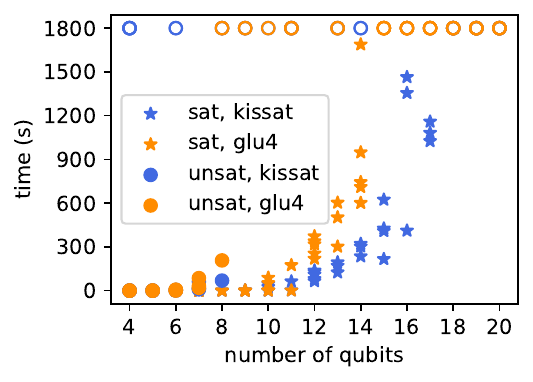}\hspace{-.15cm}%
\includegraphics[height=4.0cm]{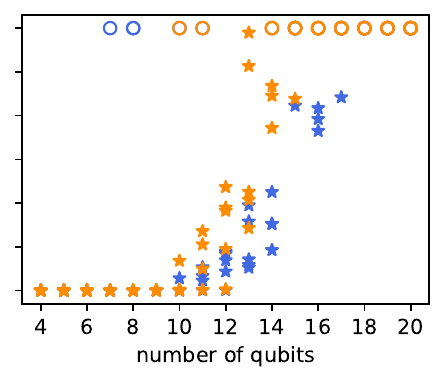}
\vspace{-.2cm}
	\caption{Synthesis under LC+VD (left) and LC+VD+limited EF (right)}
\label{fig:ghz4-time}
\end{subfigure}%
\begin{subfigure}[b]{0.3\textwidth}
\centering
\hspace{-.3cm}%
\includegraphics[height=4.0cm]{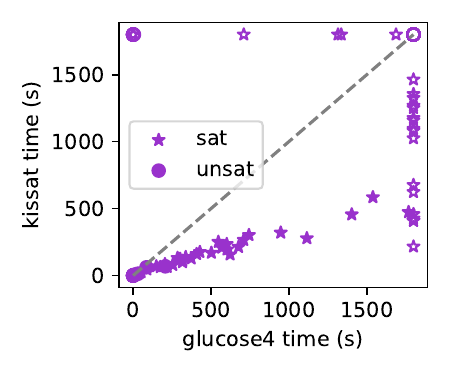}
\vspace{-.6cm}
\caption{Solver comparison}
\label{fig:solv-comp}
\end{subfigure}
\caption{The total SAT solver time for BMC with binary search over the depth up the completeness threshold (see \cref{sec:max-depth}). For each number of qubits we run on three Erd\H{o}s-R\'{e}nyi random graphs with $p = 0.8$, with a 4-qubit GHZ state as target, with only LC+VD on the left, and LC+VD+EF on a random set $D$ with $|D|=\tfrac{1}{2}|V|$ on the right. \ref{fig:solv-comp} shows the difference between the solvers for the data points from both the left and right plot in \ref{fig:ghz4-time}. Open symbols indicate timeouts. Solid spheres indicate unreachability at the depth of the completeness threshold. The largest solved instance is for 17 qubits at $d=16$, which has a formula with $\sim$2400 variables and $\sim$300,000 clauses (see above \cref{eq:R-global} for $d=1$).} 
\label{fig:bmc-time-ghz4}
\end{figure}

We first evaluate our approach in a setting where the target states are 4-qubit GHZ states (see \cref{ex:secret-sharing}), matching the target states in the empirical evaluation in~\cite{dahlberg2020how}.
GHZ states are used in a large number of applications such as quantum secret sharing~\cite{hillery1999quantum} (see also \cref{ex:secret-sharing}), anonymous transfer~\cite{christandl2005quantum} and conference key agreement~\cite{ribeiro2018fully}.
The polynomial time algorithm presented in~\cite{dahlberg2020how} can only be applied when the source graph has special properties (specifically it needs to have rank-width 1).
To evaluate our method we use Erd\H{o}s-R\'{e}nyi random graphs as source graphs, which have also been used in other work concerning graph-state synthesis~\cite{li2022photonic,lee2023graphtheoretical}. Results are shown in \cref{fig:bmc-time-ghz4}.

With a timeout of 30 minutes, Kissat can synthesize transformations under LC+VD+EF (with edge flips on a selectable subset of pairs of nodes) for graphs up to 17 qubits. Determining unreachability under LC+VD, which we do using the completeness threshold, can be done up to 8 qubits by Glucose within this timeout.
When determining reachability under LC+VD our approach performs similarly to the LC+VD+EF setting, although here it does not outperform the brute-force algorithm from Dahlberg et al.~\cite{dahlberg2018transforming}, which can find LC+VD transformations for 20 node graphs in under one minute.

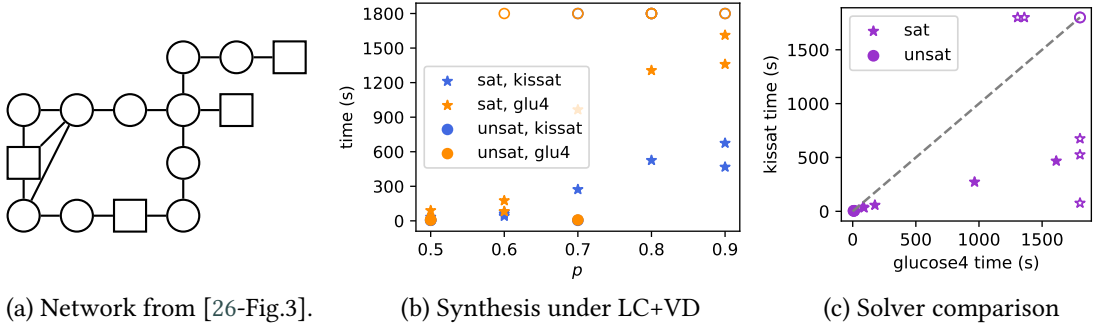
\begin{figure}[t]
\centering
\begin{subfigure}[b]{0.3\textwidth}
	\centering	
	\begin{tikzpicture}[x=7mm, y=7mm, auto, thick,node distance=.3cm,inner sep=1.5pt, end/.style={draw,minimum width=1.2em,minimum height=1.2em}, rep/.style={draw,circle,minimum width=1.2em}]
		\node[draw,end] (delft1) at (1,2) {};
		\node[draw,rep] (adam2) at (1,3) {};
		\node[draw,rep] (almere) at (2,3) {};
		\node[draw,rep] (zwolle2) at (3,3) {};
		\node[draw,rep] (zwolle1) at (4,3) {};
		\node[draw,rep] (meppel) at (4,4) {};
		\node[draw,rep] (dwingeloo) at (5,4) {};
		\node[draw,end] (groningen1) at (6,4) {};
		\node[draw,end] (enschede2) at (5,3) {};
		\node[draw,rep] (arnhem) at (4,2) {};
		\node[draw,rep] (venlo) at (4,1) {};
		\node[draw,end] (maastricht) at (3,1) {};
		\node[draw,rep] (eindhoven1) at (2,1) {};
		\node[draw,rep] (nieuwegein) at (1,1) {};
		
		\draw[-] (delft1) to (adam2);
		\draw[-] (delft1) to (almere);
		\draw[-] (delft1) to (nieuwegein);
		\draw[-] (adam2) to (almere);
		\draw[-] (almere) to (zwolle2);
		\draw[-] (zwolle2) to (zwolle1);
		\draw[-] (zwolle1) to (meppel);
		\draw[-] (meppel) to (dwingeloo);
		\draw[-] (dwingeloo) to (groningen1);
		\draw[-] (zwolle1) to (enschede2);
		\draw[-] (zwolle1) to (arnhem);
		\draw[-] (arnhem) to (venlo);
		\draw[-] (venlo) to (maastricht);
		\draw[-] (maastricht) to (eindhoven1);
		\draw[-] (eindhoven1) to (nieuwegein);
		\draw[-] (nieuwegein) to (almere);
	\end{tikzpicture}
	\vspace{.6cm}
\caption{Network from \cite[Fig.3]{rabbie2022designing}.}
\label{fig:rabbie-graph}
\end{subfigure}%
\begin{subfigure}[b]{0.4\textwidth}
\includegraphics[height=4.0cm]{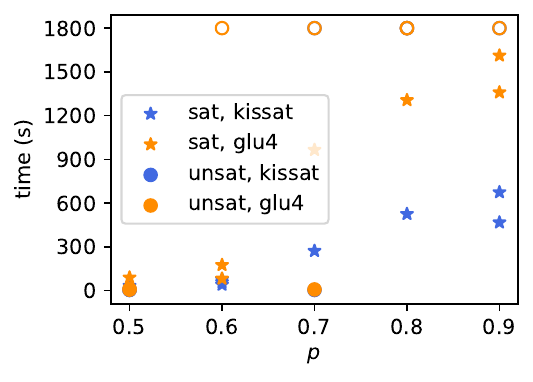}
\vspace{-.2cm}
\caption{Synthesis under LC+VD}
\label{fig:rabbie-plot}
\end{subfigure}%
\begin{subfigure}[b]{0.3\textwidth}
\hspace{-.5cm}
\includegraphics[height=4.0cm]{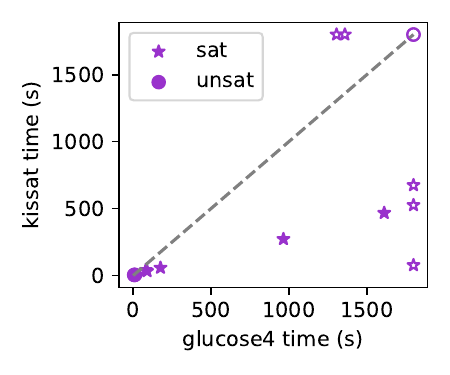}
\vspace{-.6cm}
\caption{Solver comparison}
\label{fig:rabbie-solvers}
\end{subfigure}
\caption{The 14-node quantum network proposed in~\cite[Fig.3]{rabbie2022designing}, and the SAT solver time to synthesize a transformation into a GHZ state for different amounts of entanglement ($p$) in the network. Open circles indicate timeouts. Solid spheres indicate unreachability as at the depth of the completeness threshold.}
\label{fig:rabbie-case-study}
\end{figure}

Finally, we evaluate our approach on the specific quantum network architecture proposed in \cite[Fig.3]{rabbie2022designing} (visualized in \cref{fig:rabbie-graph}). As source states, we consider graphs with nodes from this network, and random edges as follows: $(u,v) \in E$ with probability $p^{d}$, where $d$ is the distance (number of hops + 1) between the nodes, motivated by the fact that generating entanglement over larger distances is harder~\cite{sangouard2011quantum}. The target state is a GHZ state between the main network nodes (squares in \autoref{fig:rabbie-graph}). \cref{fig:rabbie-plot} shows the results for varying $p$. A higher $p$ corresponds to a larger amount of entanglement in the network. We observe that for fixed number of nodes, the time it takes to synthesize a transformation increases with the density of the source graph.

%% file: sections/cnf-appendix.tex
\interdisplaylinepenalty=10000

\section{Writing the transition relation in CNF}
\label{app:cnf}
In this appendix, we give full details on the encoding of the transition relation from \cref{sec:encoding} in conjunctive normal form (CNF), i.e. a conjunction over clauses which are each a disjunction over literals (a literal is variable $x_{uv}$ or its negation $\neg x_{uv}$). For this we use standard identities such as distributivity ($x \lor (y \land z) = (x \lor y) \land (x \lor z)$) and De Morgan's law ($\lnot (x \land y) = (\lnot x \lor \lnot y)$).

We start with the expressions for $\VDf_\vv$, $\LCf_\vv$, and $\EFf_i$ given in \cref{eq:vd-k,eq:lc-k,eq:ef-i}.
\begin{align}
\VDf_\vv &= \Land_{(\vA,\vB) \in \mathbb U} 
\begin{cases}
\neglit{x_{\vA\vB}'} & \textnormal{ if $\vA = \vv$ or $\vB = \vv$} \\
x_{\vA\vB}' \iff x_{\vA\vB} & \textnormal{ otherwise.}
\end{cases}
\tag{\ref{eq:vd-k}}
\\
\VDf_\vv^{\text{CNF}} &= \Land_{(\vA,\vB) \in E} 
\begin{cases}
\neglit{x_{\vA\vB}'} & \textnormal{ if $\vA = \vv$ or $\vB = \vv$} \\
(x_{\vA\vB}' \lor \neglit x_{\vA\vB}) \land (\neglit x_{\vA\vB}' \lor x_{\vA\vB}) & \textnormal{ otherwise.}
\end{cases}
\end{align}

\begin{align}
\LCf_\vv &= \Land_{(\vA,\vB) \in \mathbb U}
\begin{cases}
x_{\vA\vB}' \iff \lnot (( x_{\vA\vv} \land x_{\vB\vv} ) \xor \neglit x_{uv}) & \textnormal{ if $\vA \neq \vv$ and $\vB \neq \vv$} \\
x_{\vA\vB}' \iff x_{\vA\vB}& \textnormal{ otherwise.}
\end{cases}
\tag{\ref{eq:lc-k}}
\\
\LCf_\vv^{\text{CNF}} &= \Land_{(\vA,\vB) \in E}
\begin{cases}
g & \textnormal{ if $\vA \neq \vv$ and $\vB \neq \vv$} \\
(x_{\vA\vB}' \lor \neglit x_{\vA\vB}) \land (\neglit x_{\vA\vB}' \lor x_{\vA\vB})& \textnormal{ otherwise.}
\end{cases}
\end{align}
with
{
\newcommand{\pA}{x_{\vA\vv}}
\newcommand{\pB}{x_{\vB\vv}}
\newcommand{\pC}{x_{\vA\vB}'}
\newcommand{\pD}{x_{\vA\vB}}
\newcommand{\nA}{\neglit x_{\vA\vv}}
\newcommand{\nB}{\neglit x_{\vB\vv}}
\newcommand{\nC}{\neglit x_{\vA\vB}'}
\newcommand{\nD}{\neglit x_{\vA\vB}}
\begin{align*}
g = &~
(\nA \lor \nB \lor \pC \lor \pD)
\land
(\nA \lor \nB \lor \nC \lor \nD)
\land~
(\pA \lor \pC \lor \nD)
\\ &
\land
(\pB \lor \pC \lor \nD)
\land~
(\pA \lor \nC \lor \pD)
\land~
(\pB \lor \nC \lor \pD)
\end{align*}
}
\begin{align}
\EFf_i &= \Land_{(\vA,\vB) \in \mathbb U}
\begin{cases}
x_{\vA\vB}' \xor x_{\vA\vB} &\text{if $u = u_i$ and $v = v_i$} \\
x_{\vA\vB}' \iff x_{\vA\vB} &\text{otherwise}
\end{cases}
\tag{\ref{eq:ef-i}}
\\
\EFf_i^{\text{CNF}} &= \Land_{(\vA,\vB) \in \mathbb U}
\begin{cases}
(x_{\vA\vB}' \lor x_{\vA\vB}) \land (\neglit x_{\vA\vB}' \lor \neglit x_{\vA\vB}) &\text{if $u = u_i$ and $v = v_i$} \\
(x_{\vA\vB}' \lor \neglit x_{\vA\vB}) \land (\neglit x_{\vA\vB}' \lor x_{\vA\vB}) &\text{otherwise}
\end{cases}
\end{align}

Next we need expressions for $\vec y \neq b$ and $\vec y \leq b$. 
For $n$ Boolean variables $\vec y$ and an integer $0 \leq b < 2^n$, with $\text{bin}(b)$ the binary encoding of $b$ and $b_i$ the $i$-th bit of $\text{bin}(b)$, we encode $\vec y \neq b$ and $\vec y \leq b$ as in \cref{eq:neq,eq:leq}.
\begin{align}
\label{eq:neq}
f_{\neq}(\vec y, b) &= 
\Lor_{\substack{b_i \in \text{bin}(b) \\ b_i = 0}} y_i 
~~~\lor~~~
\Lor_{\substack{b_i \in \text{bin}(b) \\ b_i = 1}} \neglit y_i  \\
\label{eq:leq}
f_{\leq}(\vec y, b) &= 
\Land_{\substack{b_i \in \text{bin}(b) \\ b_i = 0}}
\bigg(
\neglit y_i \lor
\Lor_{\substack{b_j \in \text{bin}(b) \\ j > i,~b_j = 1}}
\neglit y_j
\bigg)
\end{align}

For the CNF encoding of $R_\LCf$, $R_\VDf$, and $R_\EFf$ (\cref{eq:all-lc,eq:all-vd,eq:all-ef}), the construction $A \implies B$ can be written as $(\neglit A \lor B)$. Additionally, in the following we let $c \in f$ denote a clause $c$ in a CNF expression $f$.

\begin{align}
\tag{\ref{eq:all-lc}}
R_\LCf(\vec x, \vec x') &= \Land_{\vv \in V}
\Big[ ( \vec y = \vv \land \vec z = 0 ) \implies \LCf_\vv (\vec x, \vec x') \Big]
\\
R_\LCf^{\text{CNF}}(\vec x, \vec x') &= \Land_{\vv \in V} \Land_{c \in \LCf_\vv^{\text{CNF}}} \Big( f_{\neq}(\vec y, k) \lor f_{\neq}(\vec z, 0) \lor c \Big)
\end{align}

\begin{align}
\tag{\ref{eq:all-vd}}
R_\VDf (\vec x, \vec x') &= \Land_{\vv \in V} 
\Big[ ( \vec y = \vv \land \vec z = 1 ) \implies \VDf_\vv (\vec x, \vec x') \Big] \\
R_\VDf^{\text{CNF}} (\vec x, \vec x') &= \Land_{\vv \in V} \Land_{c \in \VDf_\vv^{\text{CNF}}} \Big( f_{\neq}(\vec y, k) \lor f_{\neq}(\vec z, 1) \lor c \Big)
\end{align}

\begin{align}
\tag{\ref{eq:all-ef}}
R_\EFf (\vec x, \vec x') &= \Land_{i \in \{1, \dots, |D|\}} 
\Big[ (\vec y = i \land \vec z = 2) \implies \EFf_i(\vec x, \vec x') \Big] \\
R_\EFf^{\text{CNF}} (\vec x, \vec x') &= \Land_{i \in \{1,\dots,|D|\}} \Land_{c \in \EFf_\vv^{\text{CNF}}} \Big( f_{\neq}(\vec y, i) \lor f_{\neq}(\vec z, 2) \lor c \Big)
\end{align}

And finally, we write $R_\Idf$ and $C$ in CNF as follows.
\begin{align}
R_\Idf &= ( \vec z = 3 ) \implies \Idf (\vec x, \vec x') \\
R_\Idf^{\text{CNF}} &= \Land_{(\vA,\vB) \in \mathbb U} 
(f_{\neq}(\vec z, 3) \lor x_{\vA\vB}' \lor \neglit x_{\vA\vB}) \land 
(f_{\neq}(\vec z, 3) \lor \neglit x_{\vA\vB}' \lor x_{\vA\vB})
\end{align}

\begin{align}
C &= (\vec{y} \leq |V| + 1 \lor z = 2) \land (\vec{y} \leq |D| + 1 \lor z \neq 2) \\
C^{\text{CNF}} &= 
\Bigg(
\Land_{c \in f_{\leq}(\vec y, |V|+1)} (c \lor z_0) \land (c \lor z_1)
\Bigg)
\land
\Bigg(
\Land_{c \in f_{\leq}(\vec y, |D|+1)} (c \lor \neglit z_0 \lor \neglit z_1)
\Bigg)
\end{align}

By using the CNF expressions for $R_\LCf$, $R_\VDf$, $R_\EFf$, $R_\Idf$, both \cref{eq:R-global,eq:gs-bmc} are in CNF without any further modification.